\documentclass[reprint,
 amsmath,amssymb,
 aps,
prb,
]{revtex4-2}

\usepackage{graphicx}% Include figure files
\usepackage{dcolumn}% Align table columns on decimal point
\usepackage{bm}% bold math
\usepackage[caption=false]{subfig}
\usepackage{amssymb}
\usepackage{amsmath}
\usepackage{commath}
\usepackage{verbatim}
\usepackage{xcolor}
\usepackage{soul}

\DeclareMathOperator{\sech}{sech}

\begin{document}

\preprint{APS/123-QED}

\title{Finding Quantum Many-Body Ground States \\ with Artificial Neural Network}% Force line breaks with \\

\author{Jiaxin Wu}
\email{wu.2339@osu.edu}
 %\altaffiliation[Also at ]{Department of Physics, Ohio State University, Columbus, Ohio 43210, USA.}%Lines break automatically or can be forced with \\
\author{Wenjuan Zhang}%
\affiliation{Department of Physics, the Ohio State University, Columbus, Ohio 43210, USA. 
}%


%\affiliation{
%\affiliation{
%\affiliation{%
% This line break forced with %\textbackslash\textbackslash
%}%



\begin{abstract}
Solving ground states of quantum many-body systems has been a long-standing problem in condensed matter physics.%{\color{blue} There are several traditional numerical methods developed to . } 
Here, we propose a new unsupervised machine learning algorithm to find the ground state of a general quantum many-body system utilizing the benefits of artificial neural network. Without assuming the specific forms of the eigenvectors, this algorithm can find the eigenvectors in an unbiased way with well controlled accuracy. As examples, we apply this algorithm to 1D Ising and Heisenberg models, where the results match very well with exact diagonalization. %{\color{blue} and found that the relative error lies within $5 \%$} {\color{orange}$\leftarrow$ (*not sure if the extra sentence is necessary because we already said the accuracy is well controlled and we have not calculated the error.*) }. 

%\begin{description}
%\end{description}
\end{abstract}

\maketitle

%\tableofcontents

\section{\label{sec:intro}Introduction}
Machine learning (ML) has become a popular topic in physics, since its ingenuity and flexibility unprecedentedly allow computers to learn automatically about the underlying physics from the input data. Previous efforts have been carried out in designing various machine learning algorithms to study different physics problems, such as using Boltzmann machine to model thermodynamic observables \cite{Torlai2016}, and using artificial neural network to study two-body scattering with short-range potentials \cite{Wu2018}. In addition, in order to identify quantum phase transitions, \citet{Wetzel2017} utilize principal component analysis and variational autoencoder, meanwhile \citet{Broecker2017} proposed a method using convolutional neural networks.

In condensed matter physics, one of the main hurdles is to find the eigenstates of interacting systems with reasonable system sizes. In the presence of interactions, the dimension of the Hilbert space grows exponentially with the system size, which prevents us from solving the Hamiltonian exactly except for small systems with exact diagonalization (ED). To solve this problem, many numerical methods have been put forward, such as quantum Monte Carlo (QMC) \cite{Foulkes2001,Suzuki1993,Ceperley1995}, density matrix renormalization group (DMRG) \cite{White1992,White1993,Schollwoeck2005,Schollwoeck2011}, and so on. Everyone of them has its own advantages and constraints. For example, QMC uses stochastic sampling based on a probability that is related to the partition function of the system. It is a powerful algorithm to simulate thermodynamic properties of many-body systems, and the many-body ground states can be deduced from finite-size scaling. However, QMC suffers from the notorious ``sign problem'' rendering it unfit for some fermionic models. DMRG, on the other hand, is able to obtain accurate results for general large 1D or quasi-1D systems by keeping the most relevant components in local reduced density matrices, yet it performs poorly for higher dimensional systems. Variants of DMRG, Projected Entangled Pair States (PEPS) and multi-scale entanglement renormalization ansatz (MERA), have been proposed as solutions for the dimensional limitation.  However DMRG and its variants use power law methods to diagonalize large matricies, redering the algorithm inefficient beyond a model dependent cut off. %{\color{blue} add reference or find out the meanding of "model dependent cut-off"}.

With new tools provided by machine learning, we want to ask whether there exist new algorithms in finding the eigenstates of a general Hamiltonian. Recently, Carleo and Troyer \cite{Carleo2017} proposed to use restricted Boltzmann machine to find a variational ground state and its time evolution for a given Hamiltonian. In this paper, we address the problem by exploring an alternative unsupervised machine learning method with artificial neural network. Our goal is to find the ground-state eigenvectors of a general Hamiltonian without any assumptions of the form the eigenvectors. 

The structure of this paper is organized as the following. We describe our machine learning algorithm in Section \ref{sec:NN_structure}. To illustrate how well this method works, we apply it to find the ground states of 1D Ising model and Heisenberg model, and compare the spin-spin correlators and ground-state energy with exact diagonalization in Section \ref{sec:result}. In the end, We conclude that using this method one is able to find the ground states accurately, and discuss how this method can potentially be applied to large systems beyond ED.

\section{\label{sec:NN_structure} Neural Network Structure}

\subsection{\label{sec:intro_tradition}Traditional Deep Learning Structure}

\begin{figure*}
\includegraphics[width=0.8\linewidth]{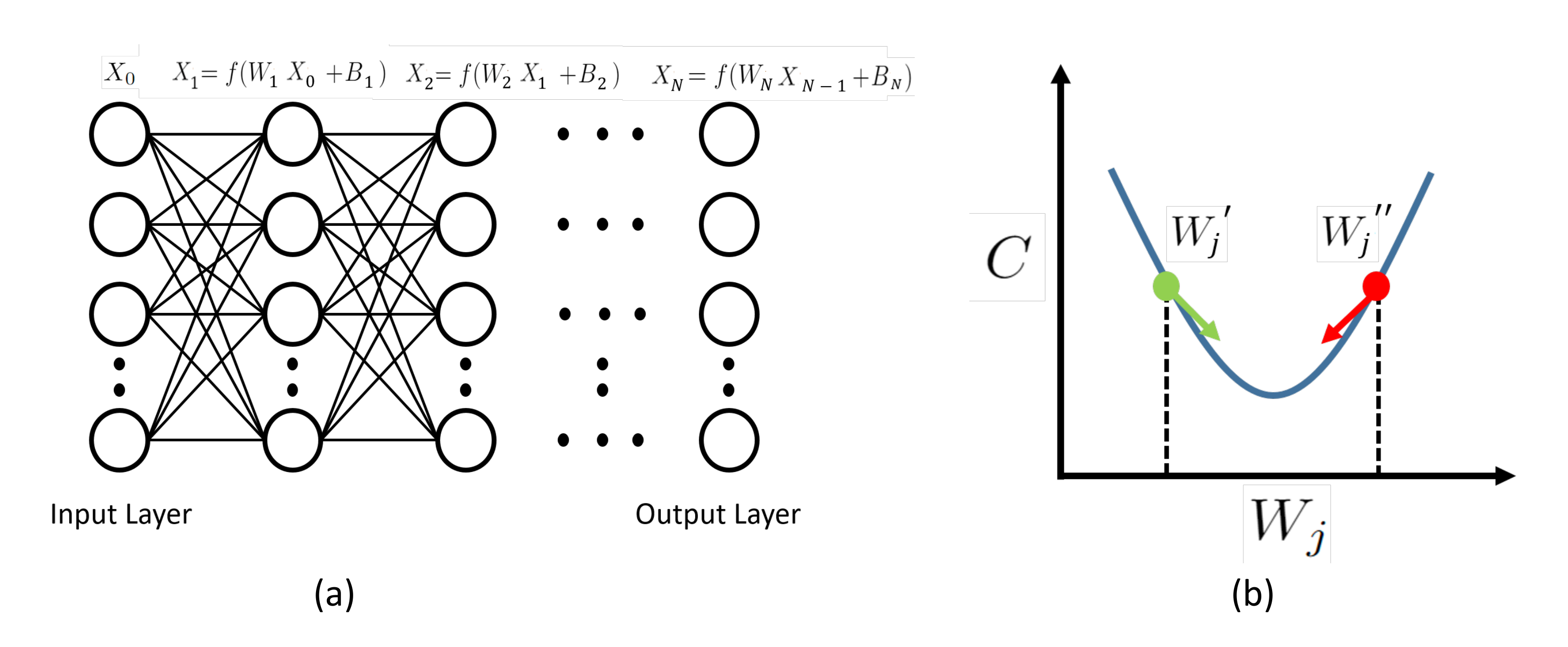}
\centering
\caption{\label{fig:nnstructure}(a) The traditional NN structure contains an input layer, some hidden layers, and an output layer. (b) Minimize the cost function $C$ in terms of $W_j$. When the partial derivative $\partial C/\partial W_j<0$, for example at $W_j'$, then the updating process increase $W_j$. If the partial derivative $\partial C/\partial W_j>0$, e.g. at $W_j'$, then the updating process decrease $W_j$. The same updating process applies to the other $W$'s and $B$'s.  }
\end{figure*}

Let us start by introducing the structure of a deep learning artificial NN. Traditionally, A deep learning NN is constructed by an input layer, multiple hidden layers, and a output layer (See Fig. \ref{fig:nnstructure}(a).). The input layer $X_0$ usually contains information of the input data, such as a vector describing every pixel of a picture. A hidden layer $X_j \ (j = 1,2,3,\cdots ,L-1)$ is a vector with arbitrary dimension, where every element is connected with every other element from the previous and next layer. The $X_j$'s are connected through parameters $W_j$'s and $B_j$'s:
\begin{equation}
    X_{j+1} = f(W_{j+1} X_j +B_{j+1}),
\end{equation}
where $W_j$ is a matrix and $B_j$ is a vector. $f$ is a non-linear function called activation function which introduces non-linearity in the model. The notation here means that the activation function is acting on every element of the vector $(W_{j+1} X_j +B_{j+1})$. The role of $f$ is important, because it is likely that $X_L$ and $X_0$ cannot be related by linear operations. Without the non-linearity from the activation function, a multi-layer NN structure has no difference from having only the input and output layers. The common choices of $f$ are sigmoid, tanh, etc. Within the same layer, the elements are independent of each other. An output layer $X_L$ can be a number or a vector containing output information from the NN. The task of a NN is to map the input $X_0$ to an output $X_L$ where $X_L$ is as close to the desired output $Y$ as possible. 

In order to accomplish the task, one needs to have a well defined cost function, which guides the direction of the learning process. It usually quantifies the difference between $X_L$ and $Y$. As one of the simplest examples, the cost function could be defined as
\begin{equation}
    C = \frac{1}{2}|Y-X_L|^2.
\end{equation}
In this case, if $X_L$ is the same as $Y$, the cost function $C=0$. The more $X_L$ deviates from $Y$, the larger $C$ is. Now the goal is to minimize $C$ by updating the parameters in the NN. Because the optimization depends on the knowledge of the desired outputs, which requires one to label the data beforehand, this is a typical supervised machine learning method. The initial parameters $W_j$'s are usually chosen randomly and the $B_j$'s are left to be zero vectors. One can optimize the output by repeatedly updating the parameters in the direction where $C$ decays the fastest. To be more precise, 
\begin{equation}\label{eq:update}
    W_j := W_j - \eta \frac{\partial C}{\partial W_j},\ B_j := B_j - \eta \frac{\partial C}{\partial B_j},
\end{equation}
where the partial derivative of $C$ is operated with respect to every matrix or vector element, and the sign ``$:=$'' denotes updating the left-hand side with the value on the right-hand side. $\eta$ is a positive number called the learning rate, and it determines how fast $C$ descends in every update. As an example shown in Fig. \ref{fig:nnstructure}(b), when $\frac{\partial C}{\partial W_j}<0 $ at point $W_j'$, the new value of $W_j$ is updated forward. While $\frac{\partial C}{\partial W_j}>0 $ at point $W_j''$, $W_j$ is updated backward. In this way, the parameters are always updated in the direction of minimizing $C$ the fastest.

For a given algorithm, one has the freedom to tune the number of layers $N$, the number of nodes in every hidden layers, the learning rate $\eta$. These are called hyper parameters. Changing these parameters can affect the performance of the algorithm. Therefore, one usually needs to have some trial runs to optimize the choice of hyper parameters.  

\begin{figure*}
\includegraphics[width = 0.9\linewidth]{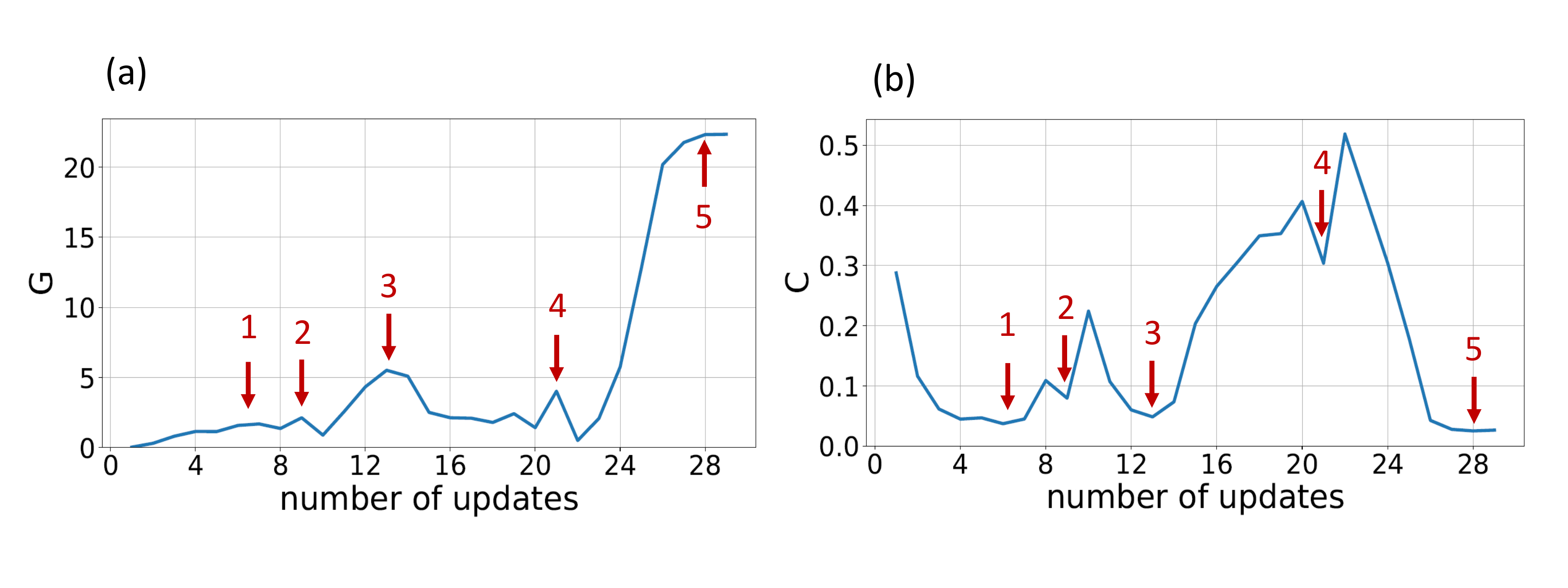}
\centering
\caption{\label{fig:G_C}(a) The typical behavior of the gain function $G$, which increases with the number of updates. We use red arrows to point out some major peaks. Only the fifth peak, which is also the highest peak, correspond to the ground state.  (b) In the same optimization process, we plot out the change of the cost function. All the peaks in $G$ have a corresponding local minimum in $C$. As $G$ increases, $C$ jumps out of four local minimums and reaches the ground state minimum indicated by arrow 5. From the cost function, it is not clear to tell whether a minimum corresponds to the ground state. The gain function, on the other hand, provides a more convincing signal, especially when one has a good estimate of the ground state energy. }
\end{figure*}

\subsection{\label{sec:modified_structure}Modified Neural Network to Find Eigenvectors }
Based on the deep learning structure, we propose a modified machine learning algorithm to find the ground state of a general Hamiltonian. Here, the input layer $X_0$ is a vector indicating the initial guess of the eigenvector given Hamiltonian $H$. Without prior knowledge of the eigenvetors, one can choose $X_0$ as a random vector. Noticed that the eigenvector for a general Hamiltonian is complex, $X_0$ is a complex vector, which can be decomposed as  
\begin{equation}
    X_0 = X_0^R + i X_0^I,
\end{equation}
where superscript $R$ and $I$ mean the real and imaginary part respectively. To simplify the calculation, we propagate the real and imaginary part of $X$ through the NN separately, i.e. 
\begin{equation}
    X_{j+1}^R = f(W_{j+1}^R X_j^R + B_{j+1}^R),\  X_{j+1}^I = f(W_{j+1}^I X_j^I + B_{j+1}^I).
\end{equation}
Here, all $W^{R/I}_j$'s and $B^{R/I}_j$'s are real, and we choose the activation function to be $\tanh$. To test whether the output vector $X_N$ is an eigenvector of $H$, we first normalize $X_N$ such that $X_N^{\dagger}\cdot X_N = 1$. Then we calculate
\begin{equation}\label{eq:HX}
    H\cdot X_N = E \tilde{X}_N,
\end{equation}
where $\tilde{X}_N$ is also properly normalized, and $E$ is a scalar. When $X_N$ is an eigenvector, $\tilde{X}_N=X_N$ and $E$ is the eigenenergy. To quantify how close $X_N$ is to be an eigenvector, we define the cost function as 
\begin{equation}\label{eq:cost}
    C \equiv 1- |\tilde{X}_N^{\dagger}\cdot X_N|^2.
\end{equation}
When $X_N$ is close to be an eigenvector, $C\rightarrow 0$; whereas when $X_N$ is far from being an eigenvector, $C \rightarrow 1$. The goal of the learning process is to minimize $C$ as close to zero as possible, which can be done with gradient descent (See appendix.\ref{appendix:grad_C}.) according to Eq.(\ref{eq:update}). 

Noticed that for the activation function $f(x)=\tanh(x)$, its derivative is the largest when $x=0$, which means the machine learns the most effectively when $W_{j+1}^{R/I} X_j^{R/I} + B_{j+1}^{R/I} \rightarrow 0$ for all $j$'s. To speed up the learning process, we initialize all elements of $W_j$'s with normal distributed random numbers multiplying by a small number (usually $10^{-2}\sim10^{-4}$), and zero all $B_j$'s. For large systems, initializing $W_j$'s with small matrix elements can greatly improve the learning speed and the accuracy of the output vectors.

The minimization process stops when $C < \epsilon_c$, where $\epsilon_c$ is the threshold for the cost function. It is a small positive number with $\epsilon_c<1$. The smaller $\epsilon_c$ is, the more accurate the output eigenstates are, and usually the more layers are required in the NN.

The cost function in Eq.(\ref{eq:cost}) has many minimums, and each of them corresponds to an eigenvector of $H$. However, the eigenvectors with an eigenenergy further away from 0 are more likely to be found. The reason is the following. Supposed that the $\{\Psi_m\}$ is the set of orthonormal eigenvectors for $H$ with eigenvalues $\{\mathcal{E}_m\}$, one can then write $X_N$ as a superposition where $X_N=\sum_m A_m \Psi_m$. In every iteration, we compute Eq.(\ref{eq:HX}), which could be written as $H\cdot \sum_m \Psi_m =E \sum_m \frac{\mathcal{E}_m}{E} \Psi_m = E \tilde{X}_N$. Under many iteration, the eigen vector correspond to the largest $\mathcal{E}_j$ is going to be more important than the other vectors by some powers of $\mathcal{E}_j$. This idea is similar to the power iteration method. In physics, we are usually most interested in the few lowest energy states. To increase the probability that the output states are one of them, one can shift the energy levels down by subtracting $H$ with a constant such that the highest energy states have eigenenergy close to 0. Therefore, after shifting the energy levels, the form of the cost function in Eq.(\ref{eq:cost}) determines that one will most likely finds the few eigenstates with the lowest eigenenergy. Since the definition of $C$ doesn't require prior knowledge of the correct eigenvectors, this is an unsupervised machine learning algorithm.

\subsection{Gain Function for Convergence to Ground States}

The cost function introduced in Section \ref{sec:modified_structure} is good when one is interested to see a few eigenstates with relatively low energy. However, in a lot of cases, we are only interested in the ground state. If $H$ has large dimensions with some low excited states, it could take quite some trials until one finds the ground state. With the cost function in Eq.(\ref{eq:cost}), one way to find it is to project out every output eigenstate in $H$ before running the next trail, and compare all the eigenvalues from these different eigenstates to determine which one corresponds to the ground state. But this projection method requires every output eigenstate to be extremely accurate, otherwise the Hamiltonian will get mixed up. In this subsection, we introduce the gain function, as opposed to the cost function, to help find the ground state more directly.

The gain function is defined as
\begin{equation}\label{eq:gain}
    G \equiv |{X}_N^{\dagger}\cdot (H X_N)|^2 = |E X_N^{\dagger} \cdot \tilde{X}_N|^2.
\end{equation}
After properly shifting the energy levels, $G$ is maximized only when $X_N$ is the ground state eigenvector. One can then use gradient ascent (See Appendix.\ref{appendix:grad_G} for more details.) 
\begin{equation}\label{eq:update_gain}
    W_j := W_j + \eta \frac{\partial G}{\partial W_j},\ B_j := B_j + \eta \frac{\partial G}{\partial B_j},
\end{equation}
to iteratively maximize $G$. In other words, the gradient ascent process refines $X_N$ to be the ground state eigenvector. 

The typical behaviors of $G$ and $C$ in the process of gradient ascent is shown in Fig. \ref{fig:G_C}. While maximizing the gain function $G$, the cost function $C$ is able to jump out of the minimums of some excited states and reach the ground state minimum. 

Unlike the cost function having a universal minimum value, the maximum value for the gain function is model dependent, which equals to the ground state energy square. Without prior knowledge, it is sometimes difficult to guess it accurately. In this case, we stop the optimization process when $G$ increase slowly and $C$ is small. To be more precise, we keep the gain function value from the last optimization step and call it $G'$. The process stops when $0<\frac{G-G'}{|G'|}<\epsilon_g$ and $C<\epsilon_c$, where $0<\epsilon_g<1$ is a small number acting as a threshold for the gain function increasing rate. The cost function is now used as a ``quality control''. It prevents the optimization process from stopping at regions where $G$ is passing by saddle points but the output vector is not an eigenvector. In general, using the gain function can significantly reduce the number of steps needed for the learning process with the same cost function threshold $\epsilon_c$. Moreover, for the same NN structure, using the gain function usually increases the accuracy of the output eigenvectors, i.e. smaller $\epsilon_c$ becomes achievable.

However, based on the stopping condition described above, there are still chances where the optimization stops before the ground state is found, namely when an excited state eigenvector also correspond to a local minimum or saddle point in $G$. This is a legitimate concern, and one should either compare several outputs with different initial condition, or judge based on estimate of the ground state energy, to decide whether the output vector is the ground state.
As shown in Fig. \ref{fig:RoG}, however, the probability of getting a ground state eigenvector is rather high when using the gain function in the optimization process.

\section{\label{sec:result}Numerical Results}

\begin{figure}
\includegraphics[width=1.0\columnwidth]{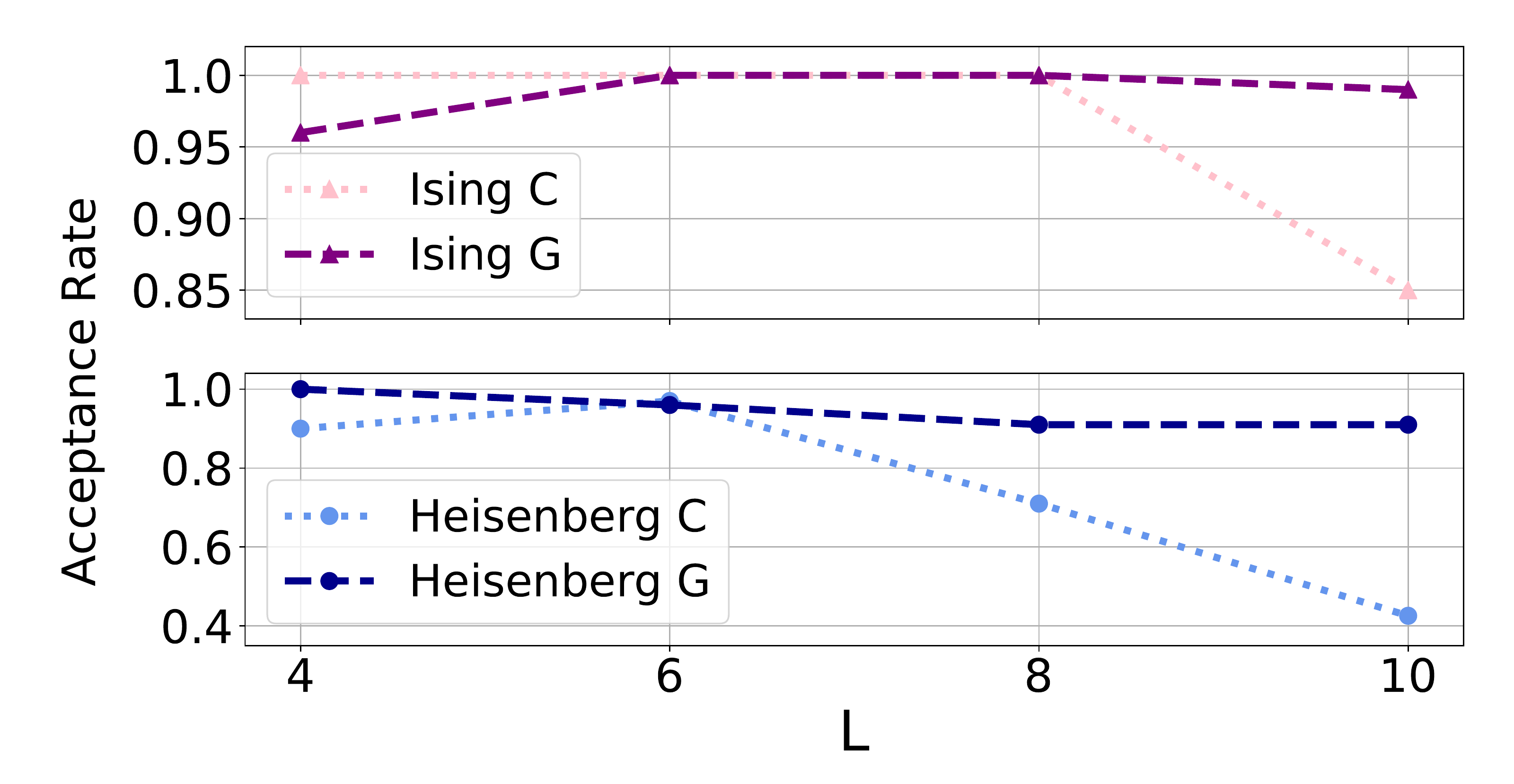}
\centering
\caption{\label{fig:RoG}The acceptance rate under different random initial condition for Ising and Heisenberg models obtained by using the cost function and the gain function respectively. With small system size, we can project the output states from the ML algorithms to the ones from exact diagonalization (ED), and determines whether the output states are ground states.}
\end{figure}
To validate the ML algorithm, we apply it to find the ground states of the 1D Ising model with staggered field and anti-ferromagnetic Heisenberg models, where the two Hamiltonian are given by
\begin{equation}
    \begin{aligned}
    H_{Ising} =& \sum_{j}[S_j^z S_{j+1}^z + h\sum_j (-1)^jS_j^z] -\frac{L-1}{4},\\
    H_{AFH} =& \sum_{j}(S^x_j S^x_{j+1}+S^y_j S^y_{j+1}+S^z_j S^z_{j+1})-\frac{L}{4},
    \end{aligned}
\end{equation}
respectively. $S_j^{\alpha} = \frac{1}{2} \sigma_j^{\alpha}$ is the spin-$\frac{1}{2}$ operator of the $j$th site with $\sigma_j^{\alpha}$'s being the Pauli matrices, and $L$ is the total number of sites in each system. In both cases, we use periodic boundary condition and shift the energy levels by adding a constant. For the Ising model, we also add a small staggered field ($h=0.1$ in our case) to split the ground state degeneracy for more straight forward comparison between the exact ground state and the output ground state. In principle, the machine learning algorithm works even if there is degeneracy, where the output ground state should be in one of the superposition of the degenerate ground states. However, the ground states found with different initial conditions are not necessarily orthogonal to each other. One way to find all the ground states is to project out the ground states from the previous calculation in the Hamiltonian before running the algorithm. However, this procedure requires every ground state is obtained with very high accuracy, otherwise the Hamiltonian will be mixed up. To avoid this complicated situation in our demonstration, we choose models without ground state degeneracy. 

\begin{figure*}
\includegraphics[height = 6.5cm,width=1.0\linewidth]{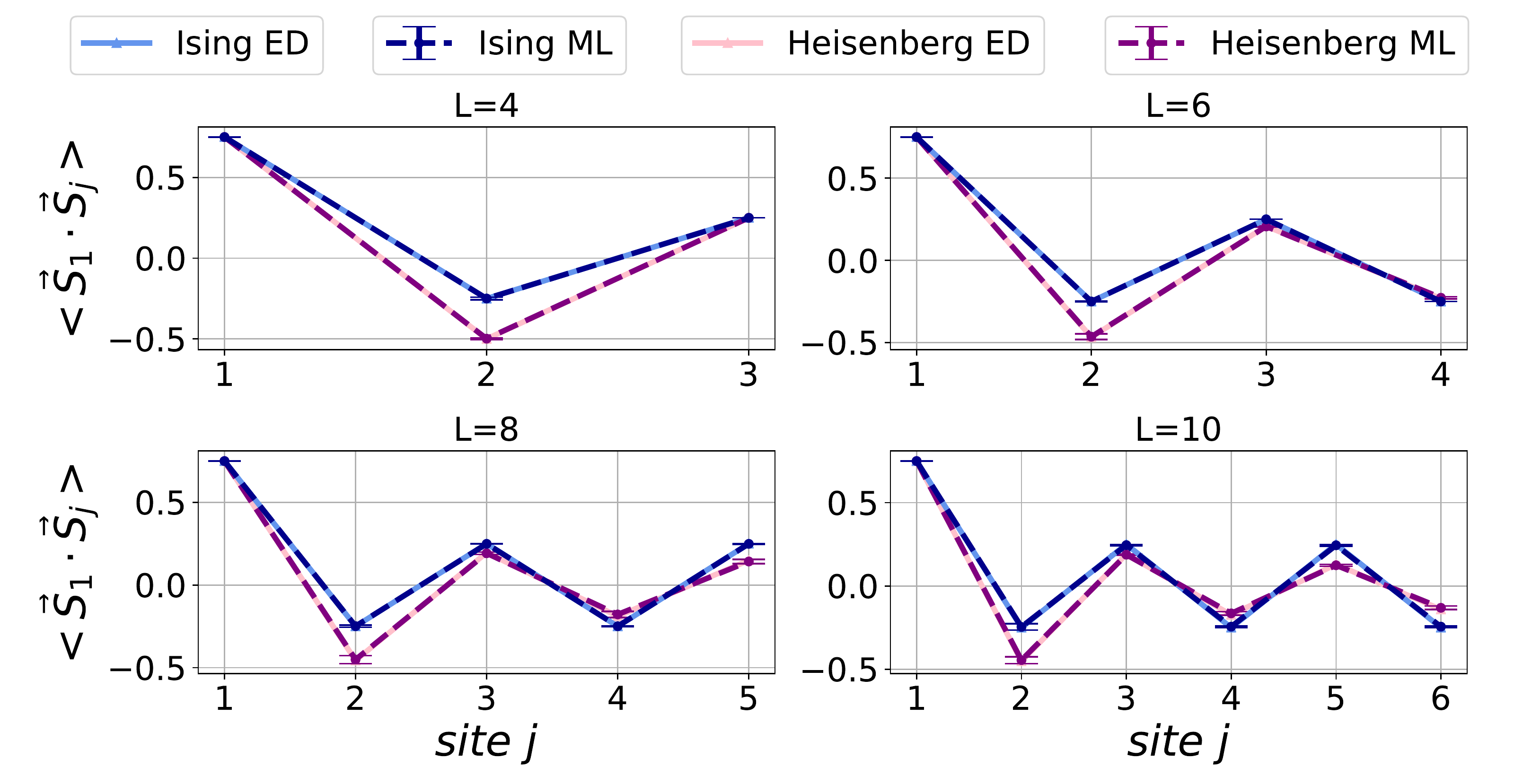}
\centering
\caption{\label{fig:corr}Two-point correlators of the first site with the other sites in the ground states with different system sizes. With periodic boundary condition, these figures cover all possible spin-spin correlators. }
\end{figure*}

In Fig. \ref{fig:RoG}, we compare the neural network's ability of finding the ground states using the cost function versus the gain function. Under many trials of random initial condition, let $N_g$ be the number of ground states found, and $N_{tot}$ be the total number of eigenstates found. We define the acceptance rate of ground states as $\frac{N_g}{N_{tot}}$. When the systems are relatively small, both methods give rather high acceptance rate. As the system size grows, the dimension of the Hilbert space grows exponentially, and it becomes less likely to find the ground states by minimizing the cost function. However, the gain function remains relatively effective as a guidance to find the ground states. Noticed that the acceptance rate also depend on the hyper parameters, so this figure only provide a qualitative ratio.

\begin{figure}
\includegraphics[width=1.0\linewidth]{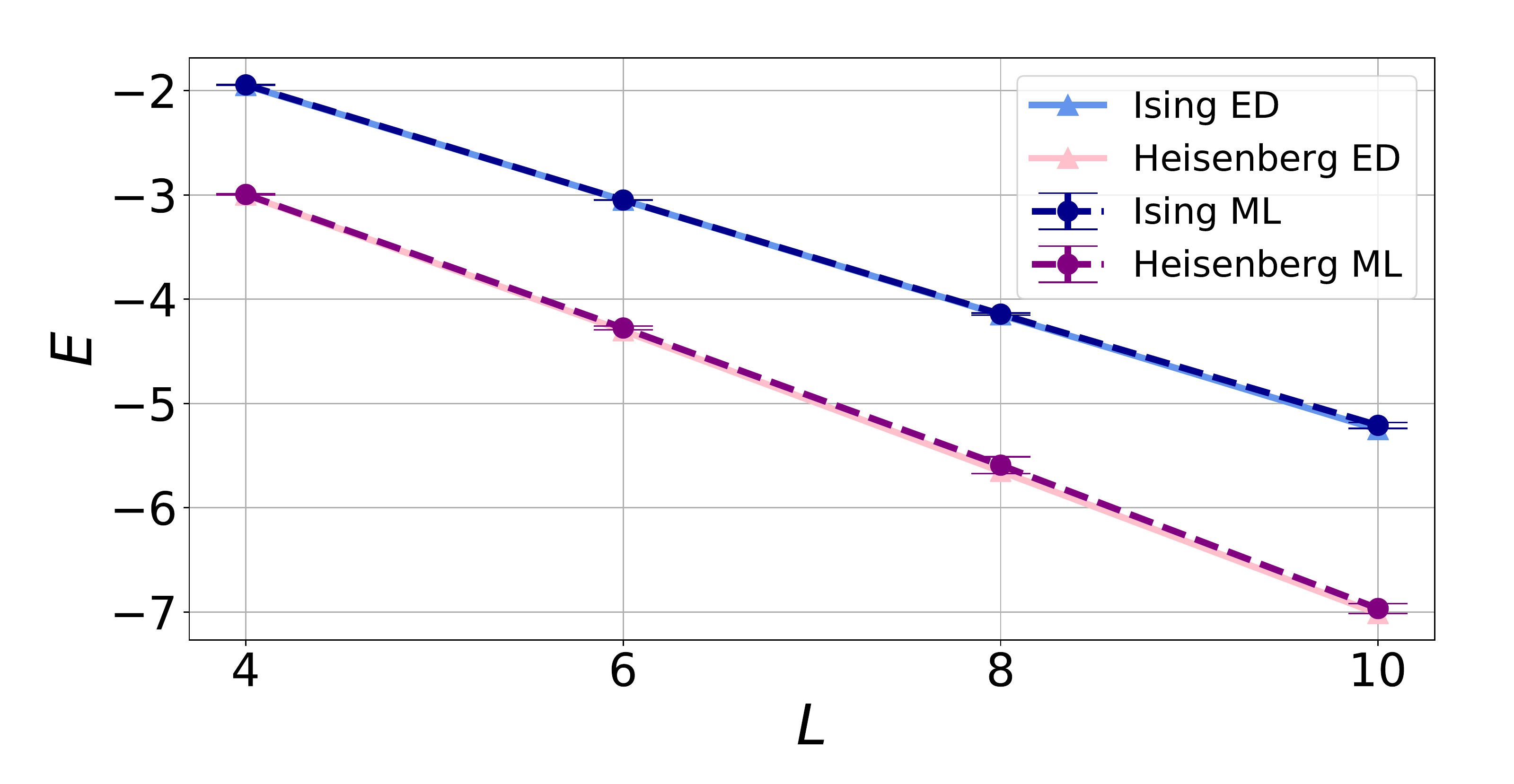}
\centering
\caption{\label{fig:EvsL}Ground state energy comparison between ED and the ML algorithm. }
\end{figure}

Next, we show the ground-state spin-spin correlations and energy calculated from maximizing the gain function versus ED in Fig. \ref{fig:corr} and Fig. \ref{fig:EvsL}. Plotted values for ML are the mean value of 100 output ground states with different random initial condition, and the error bars indicate the standard deviations in this ensemble. Fig. \ref{fig:corr} and Fig. \ref{fig:EvsL} show that the results from the ML algorithm match the ones from ED quite well.

\section{Conclusion}
In conclusion, we introduce a new machine learning algorithm to find the ground state eigenvector of a general Hamiltonian based on artificial neural network. This method does not have any constraints on the form of the Hamiltonian, nor does it require any prior knowledge of the target ground state. Moreover, the results are obtained with a controllable error rate. Therefore, the outputs are unbiased and can be made very accurate. Compare to ED, this algorithm does not involve solving any multivariable equations, but only matrix multiplications. As a result, one potential direction of applying it to large systems is to store the large matrices in hard drives and read in a few at a time for the matrix multiplication. Besides, the dimension of the matrices in the hidden layer can likely be reduced especially in the presence of symmetries in the Hamiltonian. In our discussion, we fix the dimension of the matrices in the hidden layers in order to reduce the number of hyper parameters, but there is no clear reason that this has to be the case. Reducing the dimension of these matrices can not only reduce the cost of memory, but also improve the speed of the algorithm. The minimum dimension may even reveal information about the ``order parameter'' of the Hamiltonian. Future work is needed to further explore these two possibilities.

\begin{acknowledgments}
We thank Tin-Lun Ho, Niravkumar Patel, James Rowland and Wayne Zheng for illuminating discussion. JW further acknowledges support from MURI Grant FP054294-D, the NASA Grant on Fundamental physics 1541824, and the OSU MRSEC Seed Grant. WZ acknowledges support from the National Science Foundation Grant No. DMR-1629382.
\end{acknowledgments}

\appendix
\section{\label{appendix:grad_C}Gradient Descent of Cost Function}
To update the parameters $W_j$'s and $B_j$'s, one needs to find the partial derivatives $\frac{\partial C}{\partial W_j}$ and $\frac{\partial C}{\partial B_j}$. A simple way to do it is to first calculate the partial derivatives with respect to the last layer
\begin{equation}
\frac{\partial C}{\partial B_N}=\frac{\partial C}{\partial X_N}\frac{\partial X_N}{\partial B_N}=\frac{\partial C}{\partial X_N}\frac{\partial f(Z_N)}{\partial Z_N},
\end{equation}
where $f$ is the activation function, and $Z_j = W_j X_{j-1}+B_j$. In our calculation, we choose $f(z)=\tanh(z)$. Notice that for $\tanh(z)$, its derivative is the steepest at $z=0$, which means that the machine learns the fastest when the value of $Z_j$'s are small. So when one initialize the parameters $W_j$'s, it is beneficial to choose small random numbers, and let $B_j$'s to be zero. Using the above result, one can calculate the partial derivative of the second to last layer and so on. For the $j$th layer,
\begin{equation}\label{eq:chain_rule}
    \frac{\partial C}{\partial B_j} = (\frac{\partial C}{\partial X_N}\frac{\partial X_N}{\partial X_{N-1}}\cdots\frac{\partial X_{j+1}}{\partial X_j}) \frac{\partial f(Z_j)}{\partial Z_j}.
\end{equation}
Once $\frac{\partial C}{\partial B_j}$ is known, $\frac{\partial C}{\partial W_j} = \frac{\partial C}{\partial B_j}X_{j-1}$. Repeatedly, one can update the parameters from the last layer to the first layer, and finish one updating process. This procedure is called back propagation.

With the definition of cost function C in Eq.(\ref{eq:cost}),
\begin{equation}
    \frac{\partial C}{\partial X^R_{N}}=-2 \operatorname{Re}{(\tilde{X}_N \mathcal{P})},\ \frac{\partial C}{\partial X^I_{N}}=-2 \operatorname{Im}{(\tilde{X}_N \mathcal{P})},
\end{equation}
where $\mathcal{P}$ is a scalar defined as $\mathcal{P} \equiv {\tilde{X}_N^{\dagger} \cdot X_N}$. In our algorithm, the real and imaginary parts are updated separately, i.e.
\begin{equation}
    \begin{aligned}
    \frac{\partial C}{\partial B^R_{N}} = \frac{\partial C}{\partial X^R_{N}}\frac{\partial X^R_N}{\partial B^R_{N}}=-2 \operatorname{Re}{(\tilde{X}_N \mathcal{P})} \sech^2(Z^R_{N})\\
    \frac{\partial C}{\partial B^I_{N}} = \frac{\partial C}{\partial X^I_{N}}\frac{\partial X^I_N}{\partial B^I_{N}}=-2 \operatorname{Im}{(\tilde{X}_N \mathcal{P})} \sech^2(Z^I_{N}).\\
    \end{aligned}
\end{equation}
where $Z^{R/I}_j = W^{R/I}_j X^{R/I}_{j-1}+B^{R/I}_j$, and $\sech$ is obtained from the fact that we use $\tanh$ as the activation function. For $j\in[1,N)$, we calculate $\frac{\partial C}{\partial B^{R/I}_{j}}$ based on the chain rule in Eq.(\ref{eq:chain_rule}), except that the real and imaginary parts are separate, i.e.
\begin{equation}\label{eq:chain_rule_RI}    
    \begin{aligned}
    \frac{\partial C}{\partial B^R_j} =& (\frac{\partial C}{\partial X^R_N}\frac{\partial X^R_N}{\partial X^R_{N-1}}\cdots\frac{\partial X^R_{j+1}}{\partial X^R_j}) \frac{\partial f(Z^R_j)}{\partial Z^R_j},\\
    \frac{\partial C}{\partial B^I_j} =& (\frac{\partial C}{\partial X^I_N}\frac{\partial X^I_N}{\partial X^I_{N-1}}\cdots\frac{\partial X^I_{j+1}}{\partial X^I_j}) \frac{\partial f(Z^I_j)}{\partial Z^I_j}.\\
    \end{aligned}
\end{equation}
\vspace{1em}
As for $W_j$'s, we have
\begin{equation}
    \frac{\partial C}{\partial W^R_{j}} =  \frac{\partial C}{\partial B^R_{j}} (X_{j-1}^R)^{T}, \ \frac{\partial C}{\partial W^I_{j}} =  \frac{\partial C}{\partial B^I_{j}} (X_{j-1}^I)^{T}.
\end{equation}
To minimize $C$ to zero, we update the parameters $W_j$'s and $B_j$'s using gradient descent based on Eq.(\ref{eq:update}) until $C$ is smaller than a threshold $\epsilon_{c}$, where $\epsilon_c <<1$.

\section{\label{appendix:grad_G}Gradient Ascent of Gain Function}
Similarly to gradient descent, we need to calculate the partial derivatives of $\frac{\partial G}{\partial W_j}$ and $\frac{\partial G}{\partial B_j}$ for each update for $G$'s gradient ascent. We continue to use the idea of back propagation described in Appendix.\ref{appendix:grad_C}.

For the gain function defined in Eq.(\ref{eq:gain}),
\begin{equation}
    \begin{aligned}
    \frac{\partial G}{\partial X^R_N} =& 4\operatorname{Re}(H X_N) (X_N^{\dagger} H X_N),\\ \frac{\partial G}{\partial X^I_N} =& 4\operatorname{Im}(H X_N) (X_N^{\dagger} H X_N).
    \end{aligned}
\end{equation}
therefore,
\begin{equation}
    \begin{aligned}
    \frac{\partial G}{\partial B^R_N} = \frac{\partial G}{\partial X^R_N}\frac{\partial X^R_N}{\partial B^R_N}=&4\operatorname{Re}(H X_N) (X_N^{\dagger} H X_N)\sech^2(Z^R_{N}),\\
    \frac{\partial G}{\partial B^I_N} = \frac{\partial G}{\partial X^I_N}\frac{\partial X^I_N}{\partial B^I_N}=&4\operatorname{Im}(H X_N) (X_N^{\dagger} H X_N)\sech^2(Z^I_{N})\\
    \end{aligned}
\end{equation}
For $1\leq j<N$, $\frac{\partial G}{\partial B_j}$ can be obtained through chain rule similar to Eq.(\ref{eq:chain_rule_RI}) while separating the real and imaginary parts. For $W_j$'s,
\begin{equation}
    \frac{\partial G}{\partial W^R_{j}} =  \frac{\partial G}{\partial B^R_{j}} (X_{j-1}^R)^{T}, \ \frac{\partial G}{\partial W^I_{j}} =  \frac{\partial G}{\partial B^I_{j}} (X_{j-1}^I)^{T}.
\end{equation}

Since we want to maximize $G$, we need to update the parameters in a opposite direction compared to $C$, i.e. following Eq.(\ref{eq:update_gain}) with a positive learning rate $\eta$. Furthermore, the upper bound of $G$ depends on the Hamiltonian $H$, so that we cannot define a general threshold at which the gradient ascent stops. Alternatively, we calculate the gain function from the current step $G_c$ and from the last step $G_l$, and the gradient ascent stops when the gain function almost stop increasing at consecutive steps while the cost function is under the threshold, i.e. $(G_c - G_l)/G_l < \epsilon_g$ with $(G_c - G_l)>0$ and $C<\epsilon_c$. Here, $\epsilon_g$ is the threshold for the increased ratio of the gain function and $\epsilon_g << 1$.

\newpage

\nocite{*}

\bibliography{ref_ML.bib}% Produces the bibliography via BibTeX.

\end{document}